\documentclass{aa}
\usepackage{graphics,epsfig,amsmath,amssymb,amstext,txfonts}

\begin{document}

\title{Signatures of the extragalactic cosmic-ray source composition from spectrum and shower depth measurements}

\author{D. Allard\inst{1} \and A. V. Olinto\inst{1,2} \and E. Parizot\inst{1}}

\institute{Laboratoire Astroparticule et Cosmologie (APC), Universit\'e Paris 7/CNRS, 10 rue A. Domon et L. Duquet, 75205 Paris Cedex 13, France \and The University of Chicago, 5640 S. Ellis, Chicago, IL60637, USA}

\offprints{denis.allard@apc.univ-paris7.fr}

\date{Received date; accepted date}

\abstract{We discuss the differences induced by the assumed composition of extragalactic sources on the predicted UHECR spectrum and  the energy evolution of $\langle X_{\max} \rangle$, i.e. the mean value of the atmospheric depth at the cosmic-ray air shower maximum. We show that different assumptions for the source power evolution do not modify our earlier finding that in the case of a mixed composition the ankle can be interpreted as the end of the transition from galactic to extragalactic cosmic rays. We show the characteristic features in the shape of $\langle X_{\max} \rangle$(E) that are associated with this transition for each cosmic-ray composition model. These characteristic features are present whatever the hadronic model used for the calculation. In the mixed composition cases, a signature of the interactions of nuclei with the photon backgrounds is also expected above $10^{19}$~eV. The comparisons with Stereo HiRes and Fly's Eye data favour an extragalactic mixed composition and the corresponding interpretation of the ankle. Confrontation of model predictions with future data at the highest energies will allow a better determination of the transition features and of the cosmic-ray source composition, independently of hadronic models. We also emphasize that in the pure proton case, a combined analysis of the spectrum and composition below the ankle could lead to constraints on the source power evolution with redshift.
}

\authorrunning{D. Allard,  A. V. Olinto, and E. Parizot}
\titlerunning{Signatures of the extragalactic cosmic-ray source composition}

\maketitle

\section{Introduction}
\label{Introduction}
Cosmic rays and energetic particles in general play a central role in high-energy astrophysics. In particular, ultra-high-energy cosmic rays (UHECRs) focus considerable interest not only because they are intriguing astrophysical objects by themselves, whose sources remain quite mysterious, but also because they may be helpful messengers from powerful sources in the universe, and they give access in principle to very high-energy physics inaccessible in terrestrial experiments. To identify their sources, better understand the corresponding acceleration mechanisms, and make the most of UHECRs as tools for high-energy physics and/or astrophysics, one first needs to understand their global phenomenology, which implies answering a number of pending questions, notably about the existence and shape of the expected ``GZK feature'' around $10^{20}$~eV -- i.e. the sharp decrease of the UHECR flux predicted some forty years ago by Greisen (1966) and Zatsepin and Kuzmin (1966) -- but also about the relative contribution of galactic and extragalactic sources as a function of energy.

The so-called galactic/extragalactic transition has been recognized recently as a key issue, because its actual energy range and composition structure are interrelated and contain important information about both the Galactic cosmic-ray (GCR) and extragalactic cosmic-ray (EGCR) components.

On the one hand, cosmic rays above $\sim 10^{19}$~eV are generally thought to be of extragalactic origin, notably because they seem to be dominated by proton (see e.g. Abbassi et al., 2005, and the review by Dova et al., 2005) which would either not be confined by the Galactic magnetic fields or give rise to anisotropies associated with the Galactic plane in excess of the current upper limits.
On the other hand, the Galactic origin of low-energy cosmic rays -- say below $10^{17}$~eV -- is also widely accepted. The KASCADE experiment estimated the cosmic ray flux and composition between $10^{15}$ and $10^{17}$ eV (Antoni et al., 2005), confirming the so-called \emph{knee} feature in the spectrum at $E_{\mathrm{knee}} \simeq 4\,10^{15}$ eV and showing a transition towards muon-richer cosmic-ray showers at higher energy. This trend is generally interpreted as a signature of heavier primary nuclei, independently of the assumptions relating to the underlying hadronic model used to simulate the development of CR-induced showers. The CR composition thus appears to become heavier and heavier between $E_{\mathrm{knee}}$ and $10^{17}$ eV.

Attempts have been made to determine the differential flux of individual elements (or groups of elements) among the cosmic rays in the knee region: they suggest the existence of successive knees at energies scaling with either the mass or the charge of the nuclei. However, detailed composition measurements by KASCADE are hadronic model dependent and a fully consistent picture of the measured properties of air showers in the knee region has not been reached yet. Therefore, a number of different interpretations of the knee remain compatible with the observations, including a direct signature of the maximum energy achieved by the acceleration process in Galactic sources, a change in the diffusion regime and thus of the confinement time of the cosmic rays in the interstellar medium, the local contribution of a single, relatively nearby and recent source of cosmic-rays dominating the total flux in this energy range (Erlykin and Wolfendale, 1997, 2004), or even the signature of new physics at the multi-TeV scale, where a new interaction channel towards undetected particles could be opened in the CR-induced atmospheric showers, leading to an underestimate of the primary cosmic-ray energy (Kazanas and Nicolaidis, 2001).

To properly describe the GCR/EGCR transition, a better knowledge of the composition and spectrum at the end of the GCR component would be needed. Unfortunately, even the most abundant element at the energy of the knee is not unambiguously identified (Antoni et al., 2005). A clear determination of the mass or charge scaling of the successive elemental knees would be important to discriminate between above-mentioned models. Finally, the KASCADE data indicate that the composition is still heavy at $10^{17}$~eV, but do not constrain the actual energy at which this heavy component (possibly Si or Fe group) drops out. KASCADE-Grande will significantly extend the energy range of composition analyses, but their clear interpretation will likely remain hadronic model dependent.

As a consequence of the lack of constraints on the end of the galactic component, several models accounting for the emerging extragalactic component and implying different energy ranges for the GCR/EGCR transition are currently viable. The presence of an \emph{ankle} (hardening) in the cosmic ray spectrum at $E_{ankle} \simeq 3$--6~EeV has been first interpreted as a natural signature of the transition, with a pure proton extragalactic component with a harder spectral index taking over at the ankle (equal contribution of GCRs and EGCRs), and totally dominating the spectrum above $10^{19}$~eV (see the recent study by Wibig and Wolfendale, 2004, and references therein). More recently (Allard et al., 2005a), we proposed that the source composition of the extragalactic component could be mixed (say with nuclear abundances similar to those of low-energy GCRs), in which case the extragalactic component could account for the totality of the cosmic ray flux down to the ankle, with an injection EGCR spectral index between 2.1 and 2.3 (Allard et al., 2006). In this model, the ankle appears as the signature of the \emph{end} of the transition.

On the other hand, it has been shown that an extragalactic pure-proton component with a softer injection spectrum (with an index of 2.6--2.7 assuming a uniform source distribution) can reproduce accurately the ankle feature down to $\sim 10^{18}$ eV, without any contribution from an additional Galactic component (Berezinsky et al., 2002, 2004, 2005; Aloisio et al., 2006). In this case, the ankle is the signature of the interaction of protons with CMB photons via the pair production mechanism and the GCR/EGCR transition takes place at a lower energy, around the so-called \emph{second knee} (hereafter we refer to this model as the second-knee transition model, or SKT model).

The different models of GCR/EGCR transition have very different implications for the phenomenology of UHECRs and the interpretation of the ankle, with direct impact on the inferred composition and spectral index of the supposedly power-law source spectrum of EGCRs. Clearly, the less signatures of this transition depend on hadronic models, the more robust is the determination of the underlying cosmic-ray sources. The energy evolution of the mean value of the atmospheric depth at air shower maximum, $\langle X_{\max} \rangle$, is a powerful tool to study changes in composition at the highest energies. Although the interpretation of the measured values of $\langle X_{\max} \rangle$ in terms of mass composition is hadronic model dependent, the global shape of the evolution of $\langle X_{\max} \rangle$ with energy (e.g. inflections and/or abrupt changes) is largely model-independent and can thus be directly confronted with the predictions of the different transition models.

For instance, the HiRes-MIA experiment (Abu-Zayyad, et al., 2000) reported a steep $\langle X_{\max} \rangle$ evolution between $10^{17}$~eV and $10^{18}$ eV, interpreted as a lightening of the composition. A light composition is also deduced from the HiRes stereo data above $10^{18}$~eV (Abbassi et al., 2005). These two observations are usually thought to lend some weight to SKT models.  However, in Allard et al. (2005b,c) we jointly calculated the evolution of the composition and energy spectrum of CRs in the transition region and showed that the resulting predictions in the case of the EGCR mixed-composition model are most compatible with the observed $\langle X_{\max} \rangle$ evolution above $10^{17.5}$~eV. We also showed that features in the $\langle X_{\max} \rangle$  evolution could be used as a signature of the GCR/EGCR transition along with spectral features. 

In this paper, we return to the simultaneous calculation of the extragalactic spectrum and $\langle X_{\max} \rangle$ evolution associated with different transition scenarios and discuss their distinctive signatures. In the next section, we briefly introduce the astrophysical hypotheses of our models. We present the predicted spectra and the associated $\langle X_{\max} \rangle$ evolution for different composition and source evolution hypotheses. Finally, we discuss the general validity of the different predicted features, notably against hadronic model uncertainties, and argue that $\langle X_{\max} \rangle$ measurements are currently the most efficient way to study composition features. We also identify a new signature of mixed composition scenarios at the highest energies and conclude that the new generation experiments will set very strong constraints on the transition from galactic to extragalactic cosmic rays.

\begin{figure*}[t]
\centering
\hfill\includegraphics[height=6cm]{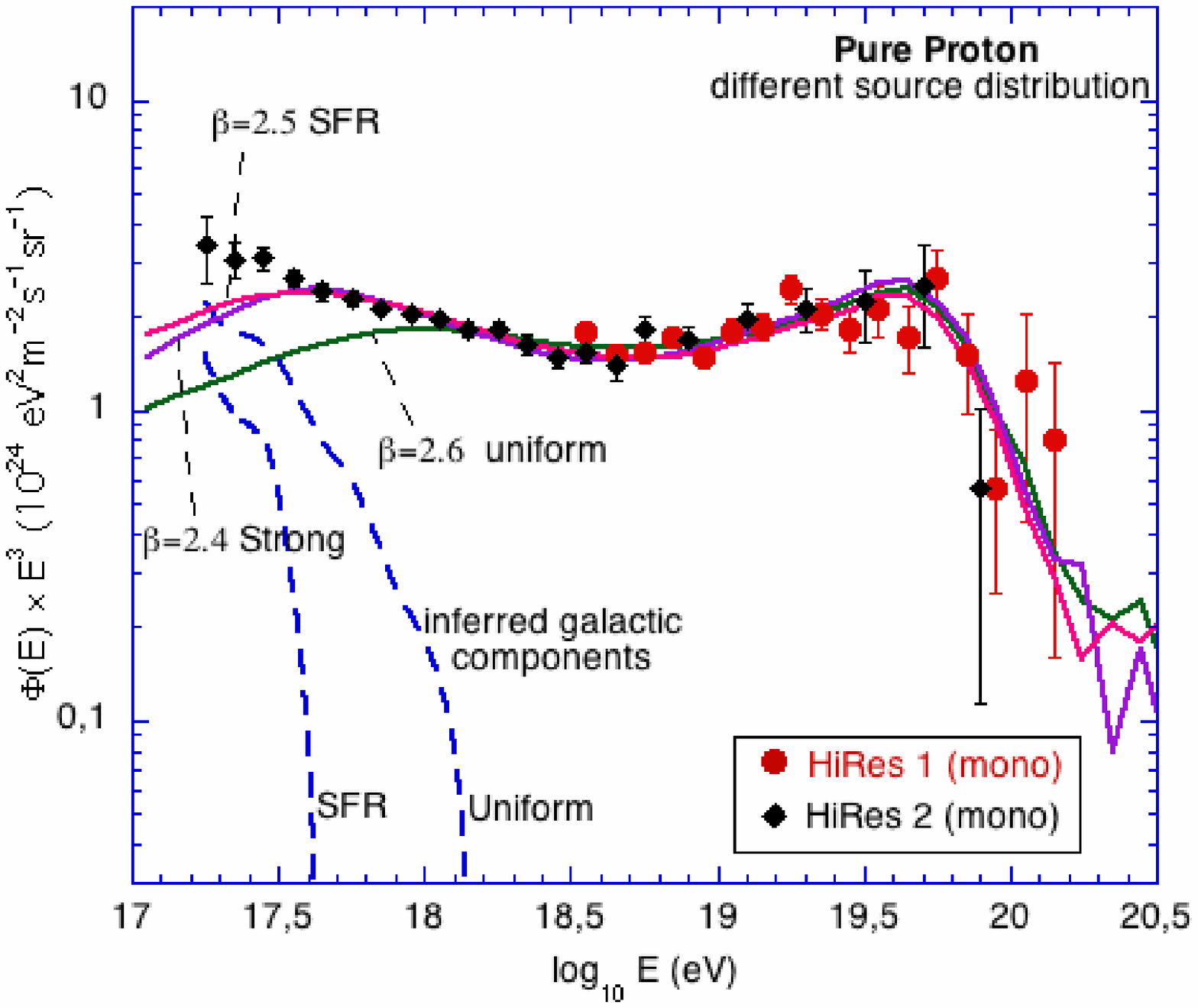}\hfill
\includegraphics[height=6cm]{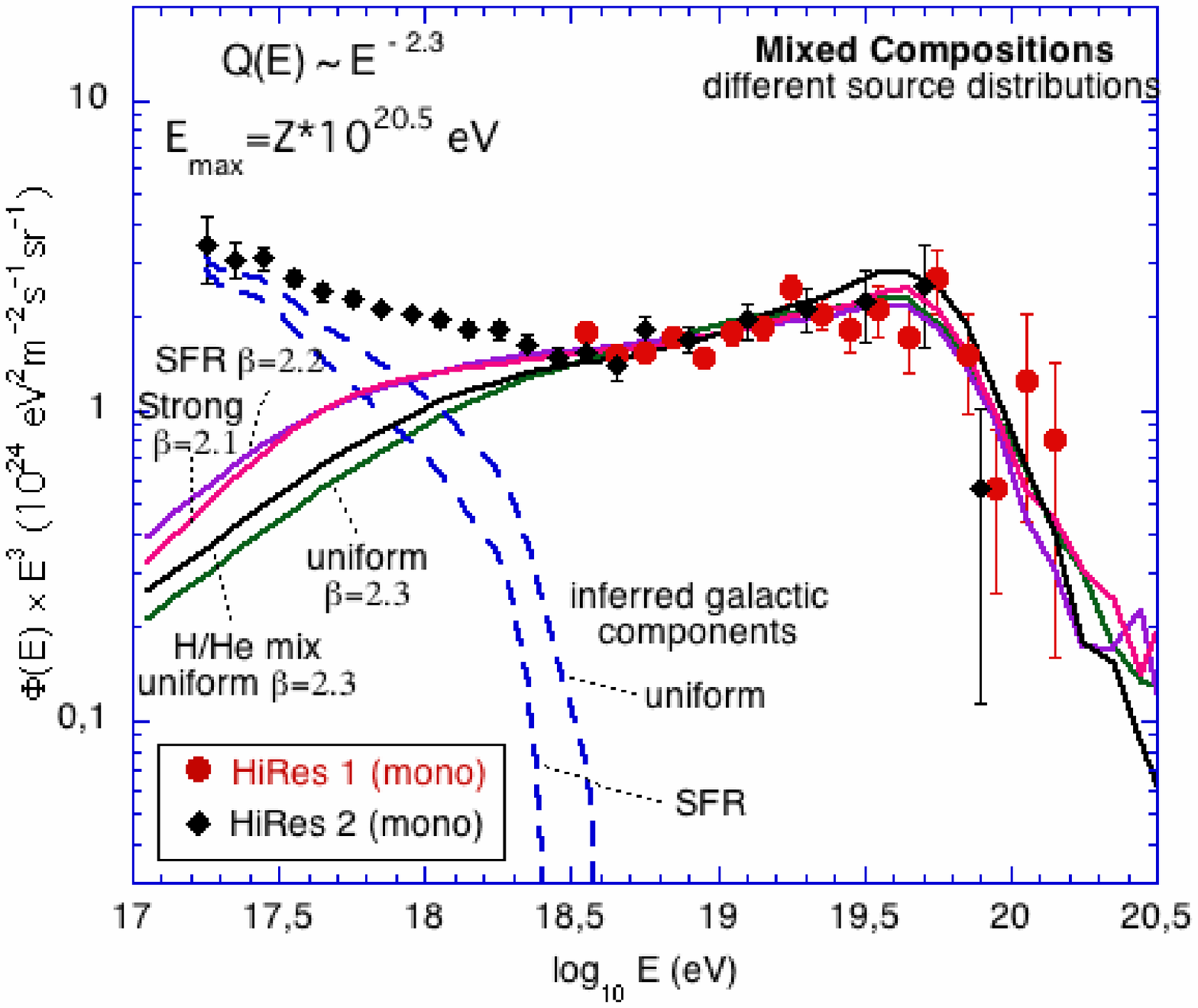}\hfill~
\caption{Propagated spectra, $E^3\Phi(E)$, for pure proton models (left) and mixed composition models (generic mix of nuclei and primordial H-He mix) (right) compared with HiRes monocular data (Bergman et al., 2005). Different source evolution models are indicated by the labels. The corresponding galactic components are inferred from the overall spectrum by subtracting the EGCR component, in the case of the uniform and SFR source evolution models (the other two cases are omitted for clarity).}
\label{Spectra}
\end{figure*}

\section{EGCR source models}

As in Allard et al. (2005a), we consider the classical pure proton scenarios and the extragalactic mixed composition models. In addition, as recalled by Hillas (2006), some specific extragalactic sources may accelerate a primordial mix of proton and helium nuclei. In our study of this case, we assume a similar H-to-He injection ratio as in the current low-energy GCR -- in other words, we scale the abundance of He nuclei in cosmic-rays by the ratio between the primordial and the current interstellar medium abundances. For our generic mixed nuclei case (Allard et al., 2005a), we assume that the EGCR source composition matches that of the GCRs observed at lower energies, and that the maximum energy achieved by nuclei of species $i$ in EGCR sources scales with their charge $Z_{i}$, i.e. $E_{\max,i}=Z_{i}  E_{\max}(^{1}H)$, as expected if the acceleration mechanism is controlled by magnetic confinement and limited by particle escape. In the following, we set the maximum proton energy to $10^{20.5}$~eV, referring to Allard et al. (2005c) for a discussion on the influence of $E_{max}$. We assume a power-law source spectrum with spectral index $\beta$, so that the number of nuclei of species $i$ (with mass number $A_{i}$) injected in the energy range $[E,E+\mathrm{d}E]$ is given by $n_{i}(E)=x_{i}A_{i}^{\beta-1}\kappa E^{-\beta}dE$ up to $E_{\max,i}$, where $\kappa$ is a normalization constant and $x_{i}$ is the corresponding abundance in the energy-per-nucleon spectrum (given by Du Vernois and Thayer, 1996).

We find that the observed UHECR spectra are best fitted with spectral indices between 2.1 and 2.3, which corresponds to a proton dominated composition with significant fractions of He and CNO, and a much lower fraction of heavier nuclei (Allard et al., 2006). However, another important ingredient of EGCR models is the time evolution of the power and/or number density of sources. Indeed, the link between the spectrum of the sources and the observed one (and thus the determination of the ``best fit spectral indices") depends strongly on the assumed redshift evolution of the sources. Here, we consider three different source evolution models. The first one corresponds to no evolution at all -- hereafter referred to as the uniform source distribution model. In the so-called ``SFR model'', we assume that the EGCR injection power is proportional to the star formation rate, as derived from the recent study of Bunker et al. (2004), which correspond a to redshift evolution in $(1+z)^3$ for $ z < 1.3$ and a constant injection rate for $1.3 < z < 6$ (with a sharp cutoff at $z = 6$). Finally, we consider a stronger source evolution favoured by the recent infra-red survey of the Spitzer telescope (Perez-Gonzalez, 2005). In this so-called  ``strong evolution model'', we assume a injection rate proportional to $(1+z)^4$ for $ z < 1$ and a constant rate for $1 < z < 6$, followed by a sharp cut-off.

\section{Propagated cosmic-ray spectra}

For each composition and source evolution hypothesis, we calculate the propagated extragalactic spectra using the latest version of our code described in detail in Allard et al. (2006). In particular, we use recent nuclear physics calculations of the giant dipolar resonance (GDR) cross sections of nuclei (Khan et al., 2005), which allow us to follow the transport of the EGCRs in the two-dimensional $(N,Z)$ nuclear space. We also use the latest results of Stecker, Malkan and Scully (2005) to estimate the intensity and redshift evolution of the infra-red, optical, and ultra-violet photon backgrounds, which play an important role in the propagation of UHE nuclei subject to photo-dissociation.

In each case, we determine the value of the spectral index $\beta$ providing the best fit of the high-energy CR data and infer the corresponding galactic component by merely subtracting the propagated EGCR component from the measured flux. The results are shown in Fig.~\ref{Spectra}. For definite predictions concerning the CR composition across the Galactic-extragalactic transition, we assume that this remaining GCR component is essentially made of iron nuclei above $10^{17.5}$~eV.

\subsection{Pure proton models}

\begin{figure}[t]
\centering{\includegraphics[height=8cm]{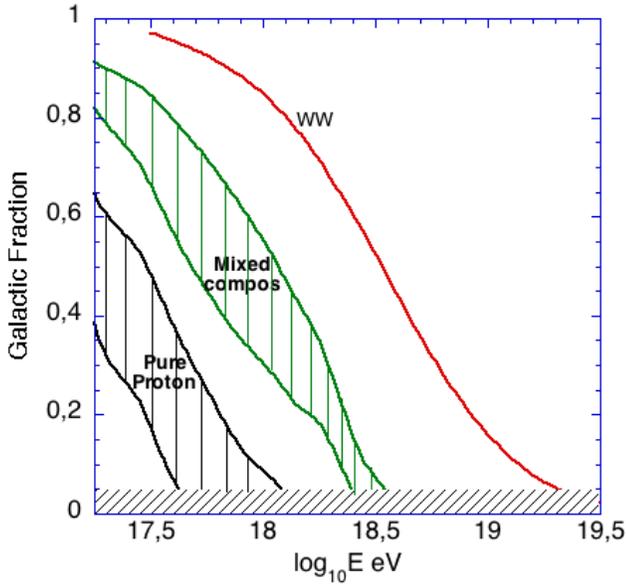}}
\caption{Fraction of the observed CRs that originate from our Galaxy as a function of energy. 
Fiducial ranges are given for the pure protons, mixed composition, and Wibig and Wolfendale (2004) (label WW)  using the HiRes 2 data for the total CR flux. The pure proton scenario would result in a steeper curve (quicker transition) in the case of a low-energy cutoff induced by extragalactic magnetic fields: the Galactic fraction would still be negligible above $10^{18}$~eV, but could match that of the mixed composition case at lower energy.}
\label{GalFrac}
\end{figure}

In the case of pure proton EGCR sources, the best fit $\beta = 2.6$ if one assumes a uniform distribution of sources (no evolution), while it goes down to 2.5 in the case of an SFR-like evolution, and 2.4 in the strong evolution case (see Fig.~\ref{Spectra}a and Berezinsky et al., 2002; De Marco et al., 2003; De Marco and Stanev, 2005; Ahlers et al. 2005; Stecker and Scully, 2005).  As shown in previous works (Berezinsky et al., 2005, De Marco and Stanev, 2005), the propagated proton spectrum and the concave shape known as the ``pair production dip'' (with a minimum around $10^{18.7}$ eV on Fig.~\ref{Spectra}a) are only mildly dependent on the source evolution hypothesis.  However, the energy where this e$^{+}$--e$^{-}$ dip begins depends on the relative weight of energy losses related to pair production, which dominate at high energy, and energy losses associated with the universal expansion, which dominate at low energy (Berezinsky et al., 2002). Therefore, the beginning of the dip depends on the redshift evolution of the source density (or power): the transition between the two energy loss processes occurs at a lower energy in the SFR and strong evolution cases (see Fig.~\ref{Spectra}). In these cases the extragalactic component can account for the whole CR flux down to much lower energies ($\sim 4\,10^{17}$~eV), which correlatively allows/requires the GCR component to cut at relatively low energies, notably lower than the confinement limit of charged nuclei in the Galaxy. Thus, in the pure proton case, the energy $E_{\mathrm{end}}$ at which the GCR/EGCR transition ends (i.e, above which cosmic-rays are purely extragalactic) depends on the source evolution scenario, as indicated by the shaded area in Fig.~\ref{GalFrac} (we discuss the possible influence of a non-negligible magnetic field below). The highest value of $E_{\mathrm{end}}$ is obtained in the case of a uniform source distribution, around 1--1.5~$10^{18}$~eV. This energy range is significantly lower than in the case of mixed composition scenarios -- a distinctive feature that can be used to discriminate between the models.

\subsection{Mixed composition models}

The propagated EGCR spectra obtained with a mixed source composition are shown in Fig. ~\ref{Spectra}b. The best fit of the high-energy data is obtained in these cases for significantly smaller spectral indices, i.e., harder source spectra: $\beta \simeq 2.3$ in the uniform case, going down to 2.2 for SFR-like source evolution and 2.1 for the strong evolution model.

\begin{table}[htdp]
\caption{Values of the EGCR source spectral index $\beta$, that best fit the UHECR data for different source compositions and source evolutions.}
\begin{center}
\begin{tabular}{c|ccc}
$\mathbf{\beta}$ & uniform & SFR & strong\\
\hline
protons only & 2.6 & 2.5 & 2.4 \\
mixed composition & 2.3 & 2.2 & 2.1 \\
\end{tabular}
\end{center}
\label{default}
\end{table}%

In all these mixed composition cases, the end of the GCR/EGCR transition roughly coincides with the ankle (Allard et al., 2005a, 2005c). Above $10^{18.5}$~eV, the predicted spectrum is quite insensitive to the source distribution, as can be seen in Fig.~\ref{Spectra}b. It is also important to note that the mixed composition models do not imply/require any definite value of the highest energy of cosmic rays in the Galactic component, as long as GCRs represent a sufficiently small fraction of the total spectrum around $E_{ankle}$ not to influence the overall spectrum and composition. Therefore, the Galactic component does not necessarily vanish above $E_{\mathrm{ankle}}$, nor is it required that cosmic rays be accelerated above $E_{\mathrm{ankle}}$ at all.

At energies below the ankle, the inferred fraction of GCRs depends on the source evolution model, just as in the pure proton case. At $10^{17.5}$~eV (resp. $10^{18}$~eV), the Galactic fraction represents  $\sim 80\%$ (resp. $\sim 50\%$) of the total flux for a uniform source distribution, and $\sim 65\%$ (resp. $\sim 30\%$) for the SFR and strong evolution hypotheses (see Fig.~\ref{GalFrac}). Thus, to avoid any misconception regarding the mixed-composition models, we stress that the energy at which the Galactic and extragalactic components have an equal contribution to the CR flux lies between $\sim 5\,10^{17}$~eV (for SFR-like or strong source evolution) and $\sim 10^{18}$~eV (for a uniform source distribution). Note also that the fraction of light elements (from the EGCR component) is a factor of two higher at $10^{17}$~eV in the strong and SFR cases than in the uniform source model. In conclusion, in our mixed composition models the cosmic rays can be dominated by light nuclei at energies below $10^{18}$~eV, which is a major difference with the GCR/EGCR transition scenario studied in Wibig and Wolfendale (2004).

Finally, we find that in the case of a primordial proton/helium mix (see Fig.~\ref{Spectra}b), the most favoured values of the source spectral indices are similar to those obtained with the generic mixed composition, for any choice of the source evolution model. No significant differences are thus found between the two types of (non pure proton) models from the point of view of the GCR/EGCR transition. We nevertheless obtain slightly lower Galactic fractions at low energies (see Hillas, 2006), because of the assumed larger proton abundance.

\begin{figure*}[ht]
\centering
\hfill\includegraphics[height=6cm]{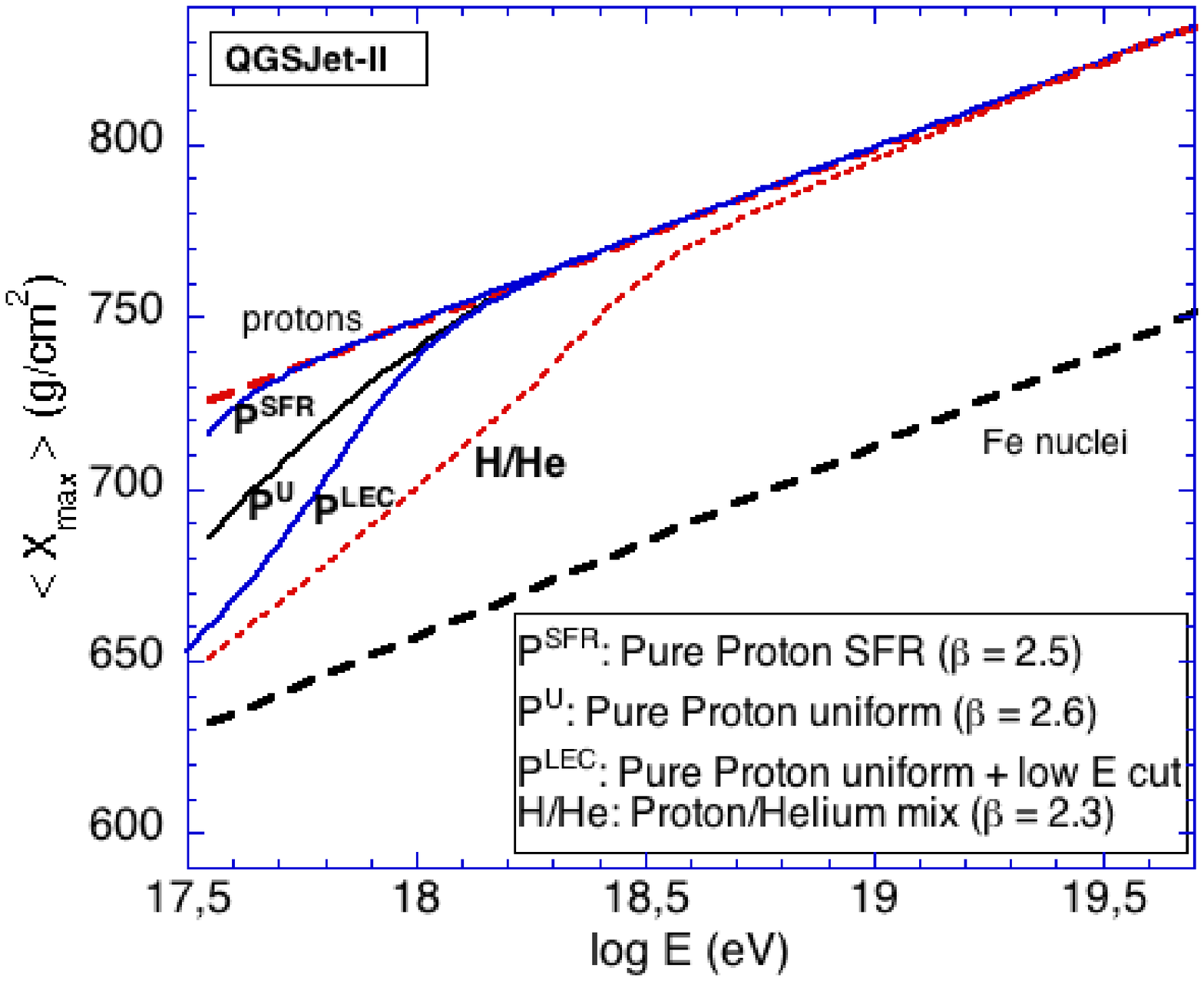}\hfill
\includegraphics[height=6cm]{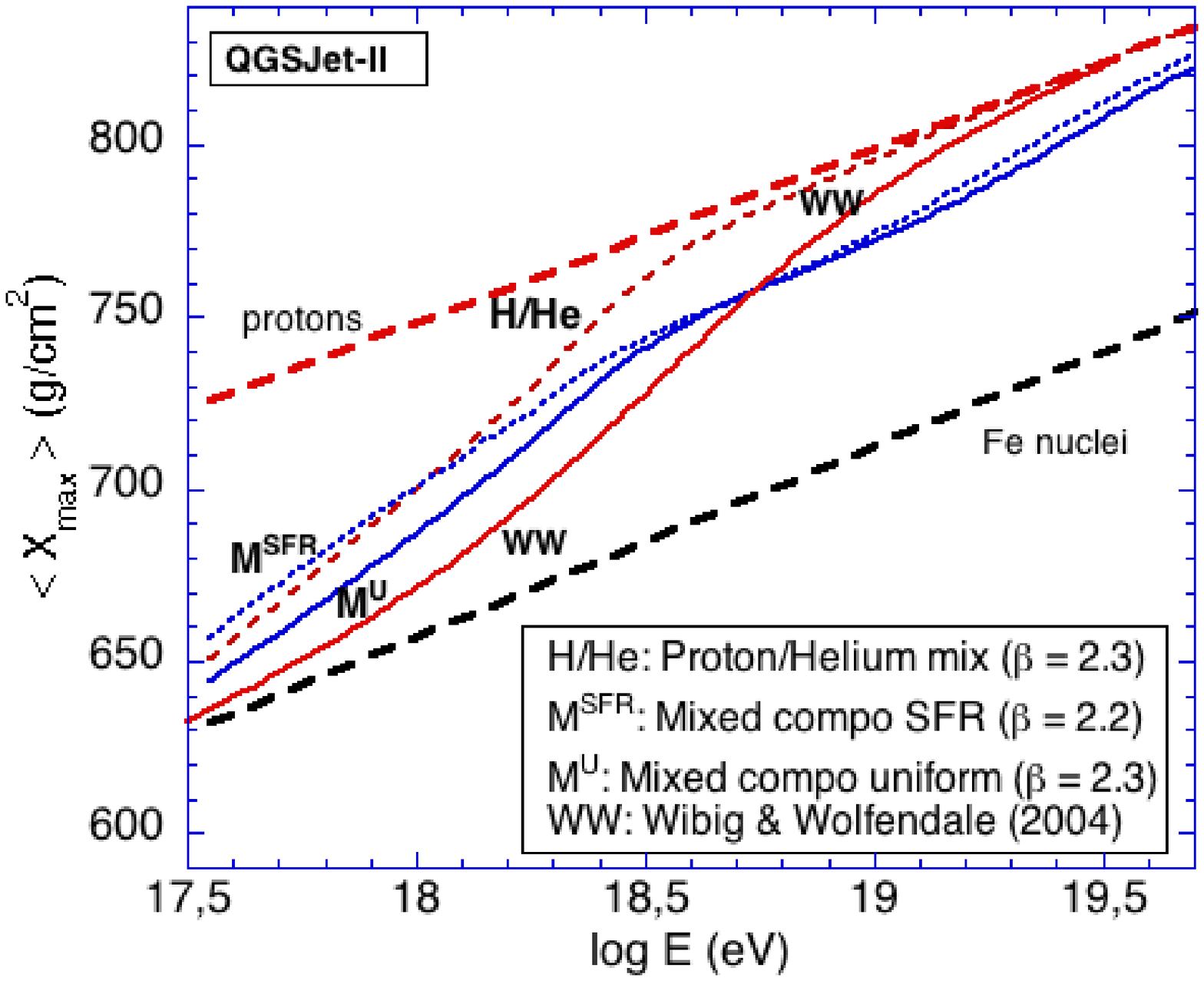}\hfill~
\caption{Evolution of $\langle X_{\max} \rangle$ as a function of energy for the different transition models (using QGSJet-II). Left: pure proton EGCR sources with two different source evolution scenarios (P$^{\mathrm{U}}$ and P$^{\mathrm{SFR}}$), and with a low-energy-cut mechanism (P$^{\mathrm{LEC}}$) as in Allard et al. (2005c). The case of a primordial mix of H and He nuclei (H/He) is also shown for comparison. Contrary to P$^{\mathrm{U}}$, P$^{\mathrm{SFR}}$ and P$^{\mathrm{LEC}}$, the latter has a GCR/EGCR transition at the ankle. Right: mixed composition EGCR sources (generic and primordial) with different source evolution scenarios. The scenario of Wibig and Wolfendale (2004) model is also displayed for comparison (WW).}
\label{Erate}
\end{figure*}

\section{The shape of $\langle X_{\max} \rangle$(E)}

Having identified the value of $\beta$ that provides the best fit of the data in each scenario, and obtained the corresponding fractions of GCRs and EGCRs at all energies, we can now deduce the evolution of the cosmic-ray composition as a function of energy and predict the values of the associated observables. The \emph{propagated} EGCR composition in each case is a direct output of our computations, and we assume that the Galactic component is essentially made of Fe nuclei above $10^{17.5}$~eV (in agreement with currently available data). From the relative abundance of all elements at a given energy, we derive the average value of the atmospheric depth (in g/cm$^{2}$) at which the maximum shower development is reached, $\langle X_{\max} \rangle$, using Monte-Carlo shower development simulations. We have used 25000 CONEX air showers and different hadronic models, namely QGSJet01 (Kalmykov and Ostapchenko, 1993), SIBYLL 2.1 (Engel et al., 1999) and QGSJet-II (Ostapchenko, 2004). Results obtained with QGSJet-II are displayed on Fig.~\ref{Erate} for the pure proton (SKT) and mixed composition models.

\begin{figure*}[t]
\centering
\hfill\includegraphics[height=6cm]{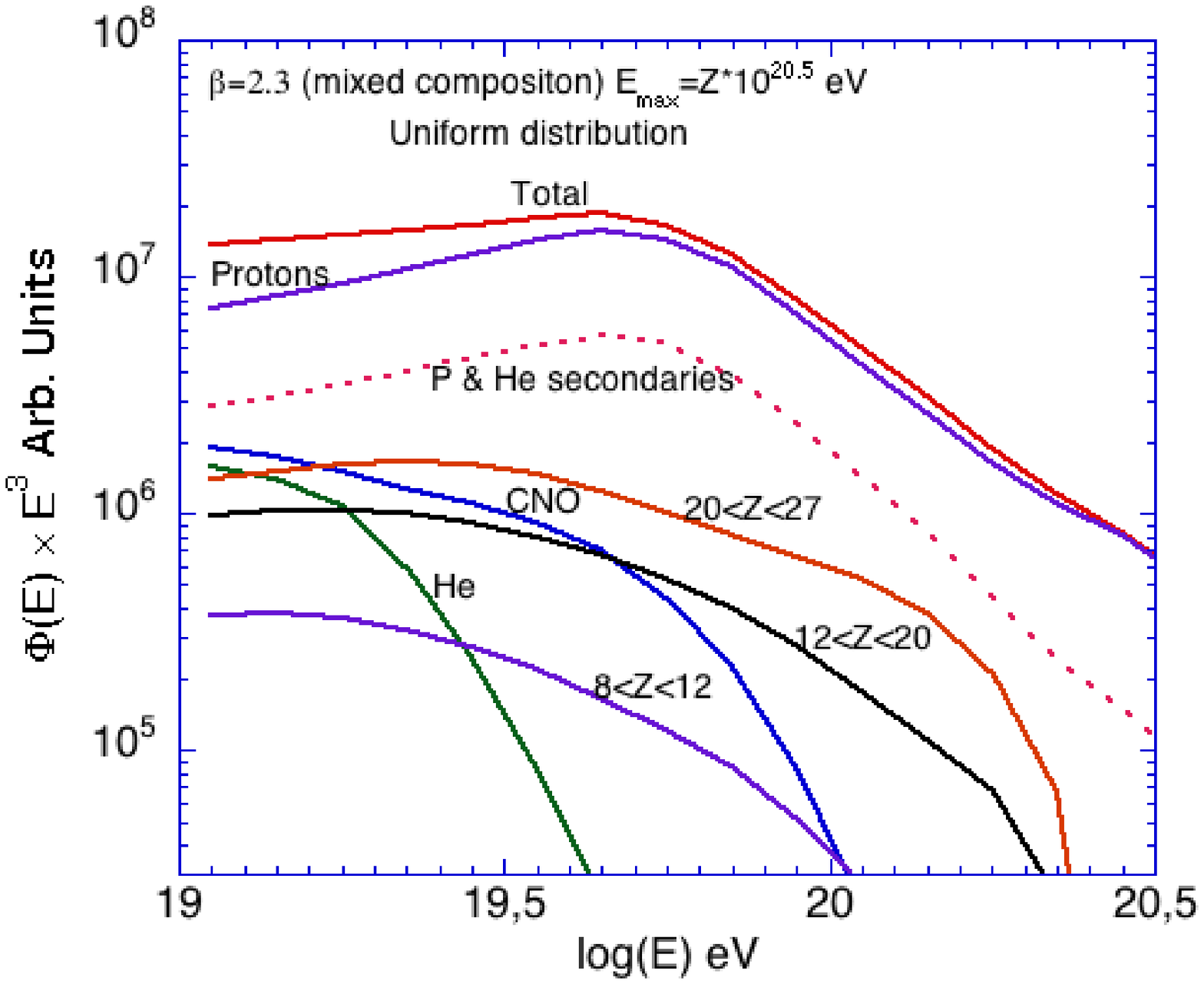}\hfill
\includegraphics[height=6cm]{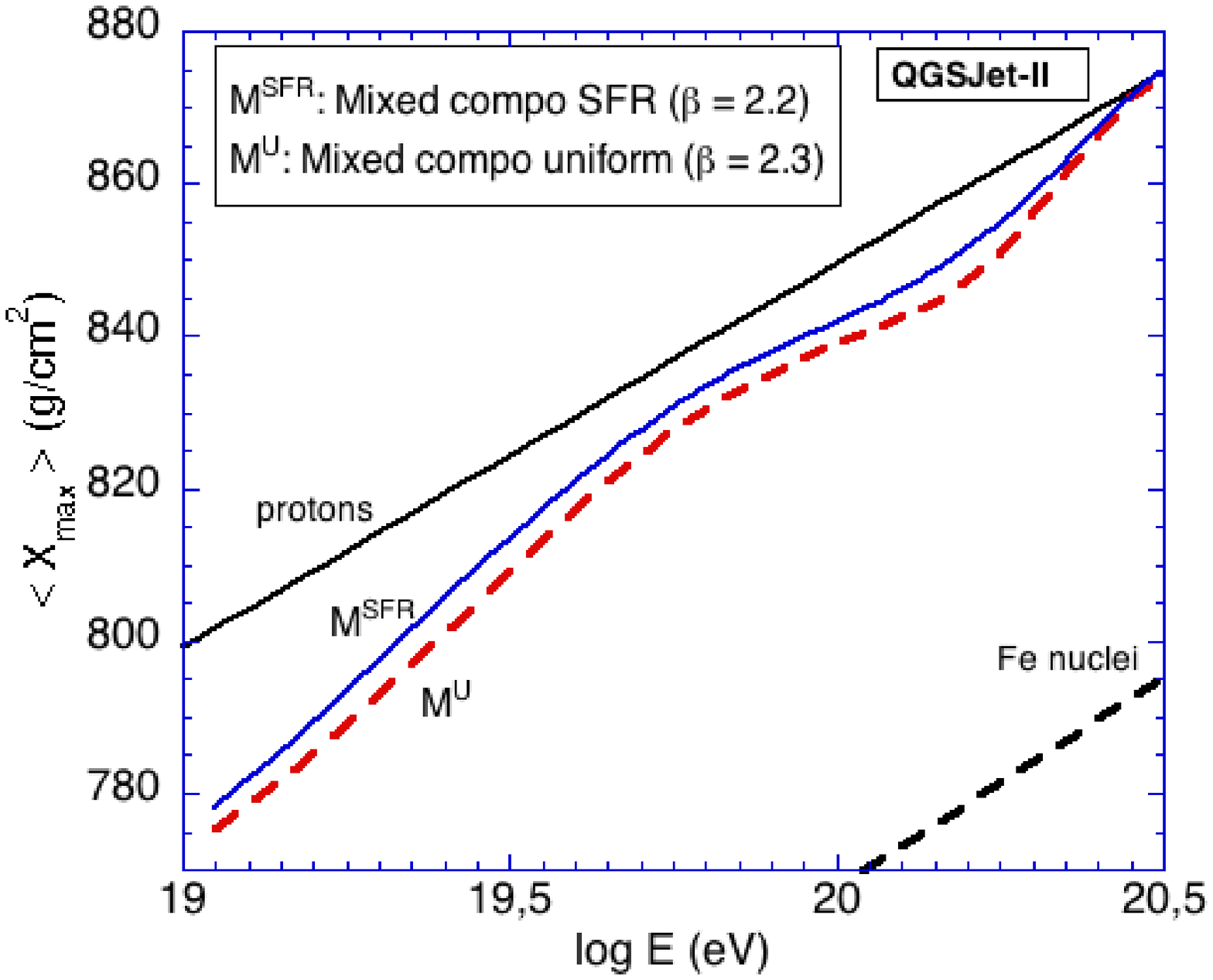}\hfill~
\caption{Left: Predicted spectrum for a mixed extragalactic composition above $10^{19}$ eV decomposed in its elemental components. Right:  $\langle X_{\max} \rangle$  evolution for an extragalactic mixed composition above  $10^{19}$ eV, for the SFR and uniform source evolution. }
\label{Erate2}
\end{figure*}

\subsection{Pure proton models}

In the pure proton case (SKT models), the interpretation of the evolution of $\langle X_{\max} \rangle$ with energy is straightforward (see Fig.~\ref{Erate}b). The transition from Galactic iron nuclei to extragalactic protons being quite narrow (i.e., it occurs over a small energy range, in a decade or even half a decade), the evolution of $\langle X_{\max} \rangle$ with energy is very steep and then gets flatter when the transition is over and the composition does not change anymore, all EGCRs being merely protons. The point where the $\langle X_{\max} \rangle$ evolution can be observed to break simply indicates the energy $E_{\mathrm{end}}$, corresponding to the end of the transition. Characteristically, an early break in the elongation rate at $\sim 4\,10^{17}$~eV is expected in the strong and SFR source evolution models (curve P$^{\mathrm{SFR}}$ on Fig.~\ref{Erate}b), whereas the break is found around 1--1.5~$10^{18}$~eV for a uniform source distribution. All pure proton models predict an observable break of $\langle X_{\max} \rangle(E)$ in the vicinity of the so-called second-knee feature, while none is expected at the ankle. Indeed, the ankle is consistently interpreted in SKT models as the signature of the interactions between EGCR protons and CMB photons producing e$^{+}$e$^{-}$ pairs. Obviously, the resulting ``pair production dip'' would not be visible if the EGCR component did not consist almost exclusively of protons. Quantitatively, nuclei heavier than H cannot contaminate the EGCR component at a higher level than $\sim 15$~\% (Wibig and Wolfendale, 2004; Berezinsky et al., 2005; Allard et al., 2005a,c; Aloisio et al. 2006).

Pure proton models may accommodate additional features, such as a break in the injection spectrum (as proposed by Berezinsky et al., 2002) or an attenuation of the EGCR flux at low energy (Lemoine, 2005; Aloisio and Berezinsky, 2005) due to a magnetic horizon effect (Deligny et al., 2004; Parizot 2004). These changes, however, would still leave the $\langle X_{\max} \rangle(E)$ evolution free from any feature at the ankle, unless the extragalactic spectrum is affected by these mechanisms at energies \emph{above} the above-mentioned predicted breaks. In this case, however, the very interpretation of the ankle as a pure pair production dip would be meaningless.

Interestingly, both cases (hereafter jointly referred-to as low-energy-cut models) lead to smaller fractions of extragalactic protons at low energy and would thus result in even steeper evolutions of $\langle X_{\max} \rangle(E)$ in the transition region. This trend is illustrated in Fig.~\ref{Erate}b, where we show the case of an injection spectrum changing from $\beta = 2.0$ below $10^{18}$~eV to $\beta = 2.6$ above (a uniform source distribution is assumed). Quantitatively, the slope of  $\langle X_{\max} \rangle(E)$ between $10^{17.5}$ and $10^{18}$~eV is 120--130~g/cm$^{2}$/decade in the standard case (depending on the assumed hadronic model) and increases up to 150--160 in the case of a low energy cut (see Fig.~\ref{Erate}b and the discussion in Allard et al., 2005c). The break point in $\langle X_{\max} \rangle(E)$, however, remains roughly at the same energy -- even though it is sharper in low-energy-cut models.

As a conclusion, the GCR/EGCR transition predicted by pure proton, SKT models is characterised by a clear break point in the evolution of $\langle X_{\max} \rangle$ with energy, located between $\sim 4\,10^{17}$~eV and 1--1.5~$10^{18}$~eV depending on the source evolution properties and the presence of additional low energy cut mechanisms.

\subsection{Mixed composition models}

The case of mixed composition models is illustrated in Fig.~\ref{Erate}b. The evolution of $\langle X_{\max} \rangle$ is relatively steep in the transition region, below $E_{ankle}$, because the composition evolves rapidly from the dominantly heavy Galactic component to the light extragalactic mixed composition. However, the evolution is significantly slower than in the case of SKT models, because the transition is wider and the cosmic-ray composition does not turn directly into protons only. As can be seen on Fig.~\ref{Erate}b, an intermediate stage appears, which may be called the mixed-composition regime, where a break in the evolution of $\langle X_{\max} \rangle$ around $E_{ankle}$ is followed by a flattening up to $\sim10^{19}$~eV, reflecting the fact that the (propagated) EGCR composition does not change much in this energy range. This is because among the different EGCR nuclei, only He nuclei interact strongly with infrared photons at these energies. Between $E_{ankle}$ and $\sim 10^{19}$~eV, the evolution of $\langle X_{\max} \rangle$ is actually compatible with what is expected from a constant composition. Then around $10^{19}$~eV, the relative abundance of nuclei heavier than protons starts to decrease significantly as a result of photo-disintegration processes: the CNO component starts interacting with the infrared background and the CMB photons eventually cause the He component to drop off completely. The evolution of $\langle X_{\max} \rangle$ therefore steepens again, accompanying the progressive evolution towards an almost pure proton composition as each type of nuclei reaches its effective (mass dependent) photo-disintegration threshold. Even though slight differences may be expected from one model to the other, the above evolution of $\langle X_{\max} \rangle(E)$ in a three steps process is a characteristic prediction of mixed-composition models, or generically of any type of EGCR sources allowing for the acceleration of a significant fraction of nuclei heavier than He (Allard et al., 2005c).

In addition to this specific signature associated with the GCR/EGCR transition, mixed-composition models can be characterised by another interesting feature appearing at the highest energies. Indeed, if Fe nuclei are accelerated above $10^{20}$~eV, the cosmic-ray composition is expected to become somewhat heavier again above $5\,10^{19}$~eV, where protons start to experience the usual GZK effect (i.e., photo-pion production over CMB photons). At this energy, heavy nuclei only interact with the infrared photons, which results in a much gentler attenuation of the heavy components than that of the protons, i.e. a heavier composition. This relative increase of the heavier component ceases around 1.5--2~$10^{20}$~eV, where interactions with the CMB photons via the GDR process take over and photo-dissociate the heavy nuclei very quickly.

From the point of view of the $\langle X_{\max} \rangle$ evolution, a second flattening is thus expected between $5\,10^{19}$~eV and $1.5\,10^{20}$~eV, followed by a final steepening towards a pure proton composition, as can be seen clearly on Fig.~\ref{Erate2}b. The actual amplitude of this feature obviously depends on the relative abundance of heavy nuclei accelerated at the highest energies, which may allow one to constrain this very important part of the EGCR injection spectrum, should such a feature be observed in the future. Note that this feature is clearly visible here in the case of our generic mixed-composition model, even though the corresponding Fe fraction at the source is less than 10\%, whatever the source evolution hypothesis and the hadronic model assumed to compute the depth of the shower maximum. While the current data above $5\,10^{19}$~eV are too scarce to test this prediction, we expect future experiments to extend composition analyses up to the highest energies, thereby helping discriminate among the EGCR source models.

\begin{figure}[t]
\centering{\includegraphics[height=8cm]{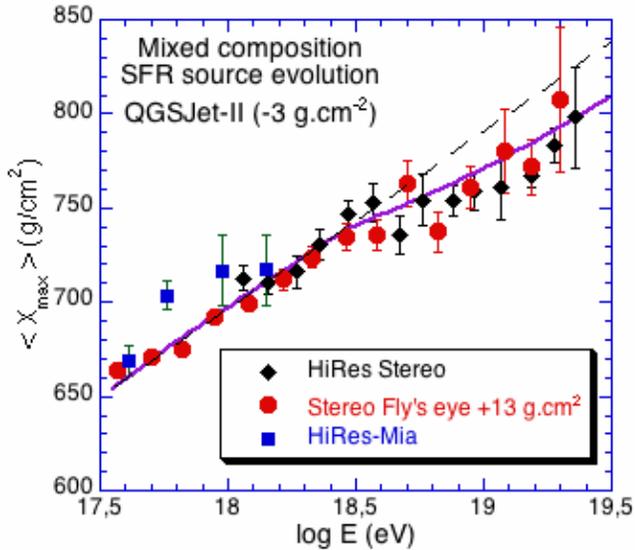}}
\caption{Predicted $\langle X_{\max} \rangle$  evolution for a mixed composition in the SFR source evolution hypothesis (QGSJet-II slightly rescaled downward by 3 $g\,cm^{-2}$) with Fly's Eye (rescaled upward by  13 $ g\,cm^{-2}$,  according to Sokolsky et al., 2005), Stereo HiRes and HiRes-Mia. }
\label{ErateComp}
\end{figure}

\section{Discussion}

The high-energy cosmic ray spectrum can be satisfactorily accounted for within either the SKT or the mixed composition models. However, we have shown that the corresponding phenomenology of the GCR/EGCR transition is very different in each case, which results in distinct shape of $\langle X_{\max} \rangle$ as a function of energy.

In both cases, we predict a break in the $\langle X_{\max} \rangle$ shape, witnessing the end of the transition from GCRs to EGCRs, but the energy at which this break occurs is characteristic of the transition model considered. In SKT models, the break  appears between $4\,10^{17}$~eV and $\sim 10^{18}$~eV, depending on the cosmological source evolution, whereas it is located in the immediate vicinity of the ankle in the mixed composition models. This difference is illustrated in Fig.~\ref{Erate}a where the $\langle X_{\max} \rangle$ evolution for a primordial mix is displayed in dashed line for comparison. In addition, in the mixed composition cases, the presence of a significant amount of nuclei heavier than He results in a steepening of the shape of $\langle X_{\max} \rangle$(E) due to photo-disintegration processes above $10^{19}$~eV. The late transition model studied in Wibig and Wolfendale (2004) is also very distinct from the point of view of the shape of $\langle X_{\max} \rangle$(E) (see Fig.~\ref{Erate}b), since the end of the transition is not expected until $10^{19}$~eV in this case. Therefore, careful measurements of the $\langle X_{\max} \rangle$ evolution at the highest cosmic-ray energies should provide an unambiguous way to discriminate between GCR/EGCR transition models (see also Allard et al., 2005c). We now argue that the $\langle X_{\max} \rangle$ signatures should be largely independent of astrophysical and physical assumptions.

In the case of a pure proton extragalactic component, it is well known that the predictions do not depend on the assumed scenario for the source power evolution, nor on the presence of reasonable extragalactic magnetic fields (see recently Aloisio et al., 2006). We also showed in this paper (see also Allard et al., 2006; Hillas, 2006) that a very good fit of the high-energy data can be obtained with an extragalactic mixed composition, whatever the source evolution model. Concerning the influence of putative extragalactic magnetic fields, we refer the reader to our discussion in Allard et al. (2005c) and simply recall here that the generic mixed composition remains dominated by protons and He nuclei. Therefore, only on small fraction of the EGCRs is likely to be affected by non negligible extragalactic magnetic fields, which suggest that our results have a large domain of validity.

From the phenomenological point of view, it is likely that extragalactic sources -- whatever their exact nature -- accelerate charged nuclei with relative abundances similar to the GCRs accelerated in standard interstellar environments. As already shown in Allard et al. (2005c), even significant modifications of the EGCR source composition do not strongly affect the above results. On the other hand, since the presence of a pair production dip in the EGCR spectrum requires a effective source composition essentially devoid of nuclei heavier than protons, SKT scenarios imply an efficient rejection of these nuclei at (or around) the source. This may be achieved by the acceleration process itself (Aloisio et al., 2006), which would then be different from the one responsible for the Galactic cosmic rays. Or it may result from a less efficient escape of heavier nuclei from sources buried in strongly magnetized structures (Sigl and Armengaud, 2005).

Concerning the predicted evolution of $\langle X_{\max} \rangle$ as a function of energy, the normalization of our results admittedly depend on the hadronic model used in the shower development simulations. It is well known that the proton-to-iron ratio extracted from the $X_{\max}$ distribution of cosmic-ray showers at a given energy are hadronic model dependent. As a matter of fact, the absolute values of $\langle X_{\max} \rangle$ as calculated using three different hadronic models (QGSJet, SIBYLL 2.1, QGSJet-II) are found to be different. However, all models predict very similar features in the shape of  $\langle X_{\max} \rangle$(E) at the very same energies (Allard et al., 2005c). This statement would also hold for other hadronic models like Nexus (Drescher et al., 2000) or the modifed version of SIBYLL 2.1 by (Drescher et al., 2005), or even the recently developed EPOS (Pierog and Werner, 2006). This is because all these models, while using different fragmentation models, predict similar trends for the shape of $\langle X_{\max} \rangle$(E), which is smooth in the whole ranges of mass and energy. However, it has been argued recently that the steep evolution of $\langle X_{\max} \rangle$ reported by HiRes-Mia between $10^{17}$ and $10^{18}$~eV may be explained not by a change of composition, but by the steepening of the $X_{\max}$ evolution caused by a modification of the hadronic interactions above $10^{17}$~eV (Alvarez-Muniz et al., 2006). It should be stressed that, although this might be true, the signatures discussed in this paper are not affected by such a model. Indeed, the characteristic feature that we propose to confront with cosmic-ray data is not the steep evolution of $\langle X_{\max} \rangle$ in the transition region itself, but the breaks in its shape, their energy scale and the possible associated features in the energy spectrum. None of the breaks we predicted could be explained in a natural way by changes in hadronic interactions, unless the energy scale of the corresponding features happen to coincide with those arising here from an astrophysical context.

Therefore, the main conclusion of this paper is that the detailed shape of the evolution of $\langle X_{\max} \rangle$ with energy, in correlation with spectral features, is a powerful way to constrain the EGCR source composition and GCR/EGCR transition phenomenology. The currently available data do not allow one to draw definitive conclusions yet. However, we argued in Allard et al. (2005c) that the predictions of the mixed-composition models appear to be in better agreement with the current data from fluorescence detectors. In particular, a good agreement is found with Fly's Eye results above $10^{17.5}$~eV (Bird et al. 1993). Concerning the slope of the $\langle X_{\max} \rangle$ evolution in the transition region (i.e., below the ankle), mixed-composition models typically predict of value between 85 and 105 g/cm$^{2}$/decade (depending on the hadronic model and the source evolution hypothesis), which is compatible with the value of 93~g/cm$^{2}$/decade reported by the HiRes-Mia experiment between $10^{17}$ and $10^{18}$ eV (Abu-Zayyad et al., 2000). Furthermore, both the predicted break at the ankle and the steepening above $10^{19}$~eV are compatible with the HiRes Stereo data (Abbasi et al., 2005; Sokolsky, 2006). Fig.~\ref{ErateComp} shows the comparison between the  $\langle X_{\max} \rangle$ evolution for a mixed composition in the SFR evolution case and the data of HiRes Stereo, HiRes-Mia and Fly's Eye (rescaled by 13 $g\,cm^{-2}$, as suggested by Sokolsky, 2005). As can be seen, Fly's Eye and Stereo HiRes data are consistent with the predicted break in the $\langle X_{\max} \rangle$ evolution between 3 and 4~EeV, which also corresponds to the energy of the ankle reported by both experiments. Note that HiRes-Mia results at lower energy are also compatible with SKT models, as well as with mixed-composition models except at one point around $5\,10^{17}$~eV.

The data best agree with the absolute scale of $\langle X_{\max} \rangle$ computed with the QGSJet-II model. However, it is important to note that the results obtained with QGSJet01 show exactly the same features shifted downwards by $\sim$20 $g\,cm^{-2}$, and are still well within the systematic uncertainties of the different experiments. This illustrates once more that the choice of the hadronic model is not critical in the present discussion.

Finally, we note that signatures of the GCR to EGCR transition can also be observed in the shape of the muon density evolution with energy in ground array observations. Again, the shape of the evolution with energy is a better signature than the traditionally quoted proton to iron ratio deduced from comparisons with shower simulations for two main reasons. First, proton to iron ratios deduced from muon densities are heavily hadronic model dependent (see, e.g., AGASA results above $10^{19}$~eV in Shinozaki et al., 2005). Second, the accuracy of the energy scale is critical for this kind of estimate as muon densities evolve strongly  with the energy ($\propto E^{0.9-0.95}$). As the difference between muon numbers  for proton and iron  is  generally $\sim$30-40$\%$  at a given energy, systematic errors in the energy estimate can easily spoil comparisons with simulations. For instance, the systematic energy shift needed to reconcile Yakutsk with Fluorescence experiments is  $\sim40\%$. Such a downward shift significantly changes  the proton to iron ratios deduced from comparisons with hadronic models  towards a much heavier composition. On the other hand, the shape of the muon density evolution with energy  appears flatter than hadronic models predictions  (Knurenko et al., 2004), which  indicates  that the composition is getting lighter at least up to the measured $E_{ankle}$ for Yakutsk. 
Whatever the real energy scale may be, the shape of the muon number evolution is qualitatively compatible with what is observed by Akeno (Dawson et al., 1998) in the same energy range (as measured relatively to the ankle) and seems to indicate that the composition is still evolving around the ankle energy (indicating that the transition may not be completed), although the statistics is not large enough to clearly identify the position of the break marking the end of the composition change (if any -- see the above discussion). In sum, with higher statistics the shape of the muon density evolution with the energy can be used to search for composition features in the same spirit as the evolution of  $\langle X_{\max} \rangle$.

Although the agreement between the experimental data and the predictions of the mixed-composition models is encouraging, more precise and higher statistics measurements are needed. In the near future, the Pierre Auger Observatory should provide accurate measurements of the energy evolution of $\langle X_{\max} \rangle$ with a lower energy threshold than HiRes Stereo, allowing for a direct test of the features predicted here. Signatures of the source evolution models may also be accessible, for instance if a break of $\langle X_{\max} \rangle(E)$ is observed around $4\,10^{17}$~eV (see Fig.~\ref{Erate}a). Interestingly, planned enhancement of the Pierre Auger Observatory using higher elevation fluorescence telescopes, in-filled water tanks and muon detectors (Medina et al., 2006, and references therein), as well as the TA/TALE experiment (Martens et al., 2006) should extend the spectrum and $\langle X_{\max} \rangle$ measurements with new generation instruments down to $10^{17}$~eV. This should allow us to investigate the beginning of the GCR/EGCR transition, the second knee, and the expected associated features in the shape of $\langle X_{\max}\rangle$(E), which are key to the detailed understanding of high-energy cosmic rays. 
 
\section*{Acknowledgments}
We wish to thank B.L.D., Michael Unger, Ralf Engel and Tanguy Pierog for helpful comments and discussions.

\end{document}